\documentclass[prl, aps, twocolumn, superscriptaddress, nobibnotes, sort&compress, reprint]{revtex4-1}

\usepackage[latin1]{inputenc}
\usepackage[USenglish]{babel}
\usepackage[pdftex]{graphicx}
\usepackage{textcomp, lmodern, xcolor, siunitx, amssymb, mathtools}
\DeclarePairedDelimiter\robra{\lparen}{\rparen} 
\DeclarePairedDelimiter\abs{\lvert}{\rvert}
\newcommand{\fig}{Fig.}

\begin{document}
\sisetup{detect-all}

\title{Centimeter-scale suspended photonic crystal mirrors}\thanks{This work was published in Opt.\ Express \textbf{26}, 1895--1909 (2018).}

\author{Jo\~{a}o P.\ Moura}\thanks{These authors contributed equally to this work.}
\author{Richard A.\ Norte}
\email{r.a.norte@tudelft.nl}
\author{\hspace{-.1cm}\normalfont\textsuperscript{,\,$\dag$} Jingkun Guo}
\author{Clemens Sch\"afermeier}
\author{Simon Gr\"oblacher}\email{s.groeblacher@tudelft.nl}
\affiliation{Kavli Institute of Nanoscience, Delft University of Technology, Lorentzweg 1, 2628CJ Delft, The Netherlands}


\begin{abstract}
Demand for lightweight, highly reflective and mechanically compliant mirrors for optics experiments has seen a significant surge. In this aspect, photonic crystal (PhC) membranes are ideal alternatives to conventional mirrors, as they provide high reflectivity with only a single suspended layer of patterned dielectric material. However, due to limitations in nanofabrication, these devices are usually not wider than \SI{300}{\micro m}. Here we experimentally demonstrate suspended PhC mirrors spanning areas up to \SI{10 x 10}{\mm}. We overcome limitations imposed by the size of the PhC and measure reflectivities greater than \SI{90}{\percent} on \SI{56}{nm} thick mirrors at a wavelength of \SI{1550}{nm} -- an unrivaled performance compared to PhC mirrors with micro scale diameters. These structures bridge the gap between nano scale technologies and macroscopic optical elements.
\end{abstract}

\maketitle

Photonic crystal (PhC) membranes are suspended dielectric sheets patterned with sub-wavelength, low-index two-dimensional periodic structures~\cite{Fan2002}. These patterns give rise to resonances that couple out-of-plane radiation to in-plane leaky modes, and can be engineered to transform a flat membrane into a mirror~\cite{Kanskar1997}, a lens~\cite{Lu2010}, or even a curved mirror~\cite{Fattal2010,Lu2010,Guo2017}. Here we study a PhC consisting of a periodic lattice of holes in a membrane, whose hole radius and lattice constant can be tuned to reflect light at a wavelength of choice. When fabricated from materials with low optical absorption such as low-pressure chemically vapor-deposited silicon nitride (LPCVD SiN), one can realize mirrors with sub-wavelength thicknesses and reflectivities $>\SI{99}{\percent}$, mostly limited by scattering losses, as shown in~\cite{Chen2017}. LPCVD SiN thin films also enable the combination of PhC mirrors with low thermal noise mechanical oscillators, due to their high intrinsic stress, thin geometry, and weak coupling to undesired thermal modes~\cite{Norte2016,Reinhardt2016}.

Microfabrication processes have so far restricted suspended PhC mirrors to areas around \SI{300 x 300}{\um}~\cite{Chen2017}. This size sets an upper bound to the waist of incident Gaussian beams, since wider waists do not completely interact with the PhC, resulting in decreased reflectivity. But the waists also have a lower bound:\ very small waists have a high divergence and couple to undesired PhC modes, which leads to shifting, broadening and shallowing of the high-reflectivity crystal resonance. These adverse finite-size effects have been consistently measured in very thin mirrors with thicknesses below $0.1\lambda$, where $\lambda$ is the wavelength of the reflected light~\cite{Bernard2016,Kemiktarak2012,Norte2016,Bui2012}. The ability to fabricate larger PhC mirrors with increasingly thinner membranes could greatly facilitate the combination of high reflectivity and low mechanical losses~\cite{Norte2016}. These properties indicate the potential that PhC mirrors may have for reducing thermal mirror coating noise which stands as a limit in precision measurements such as atomic clocks~\cite{Campbell2017}, frequency-stabilized lasers~\cite{Kessler2012}, and gravitational wave detectors~\cite{Reid2016}. At the centimeter scale, PhC mirrors could have more immediate applications as deformable mirrors with adjustable wavefront~\cite{Madec2012}, or evanescent field sensors with a large interaction area~\cite{Schmid2009,Fan2009}.

In this letter we experimentally demonstrate free-standing SiN photonic crystal mirrors with thicknesses of \num{56} and \SI{210}{nm} and areas of up to \SI{10 x 10}{mm}. Not only do we increase the area of suspended PhC mirrors by nearly 4 orders of magnitude compared to previous works, we also show that these large aspect-ratios allow us to achieve high reflectivity from membranes thinner than previously measured. We observe greater than \SI{90}{\percent} reflectivity of \SI{1550}{nm} light from mirrors with a thickness of $0.038\lambda$ (\SI{56}{nm}) -- a significant increase compared to previous devices with similar thickness and wavelength~\cite{Bernard2016}. Such large structures allow studying the spectrum of PhC membranes as a function of incident beam waist with less constraints from finite size effects.

\begin{figure}[t]
	\centering
	\includegraphics[width = 1\columnwidth]{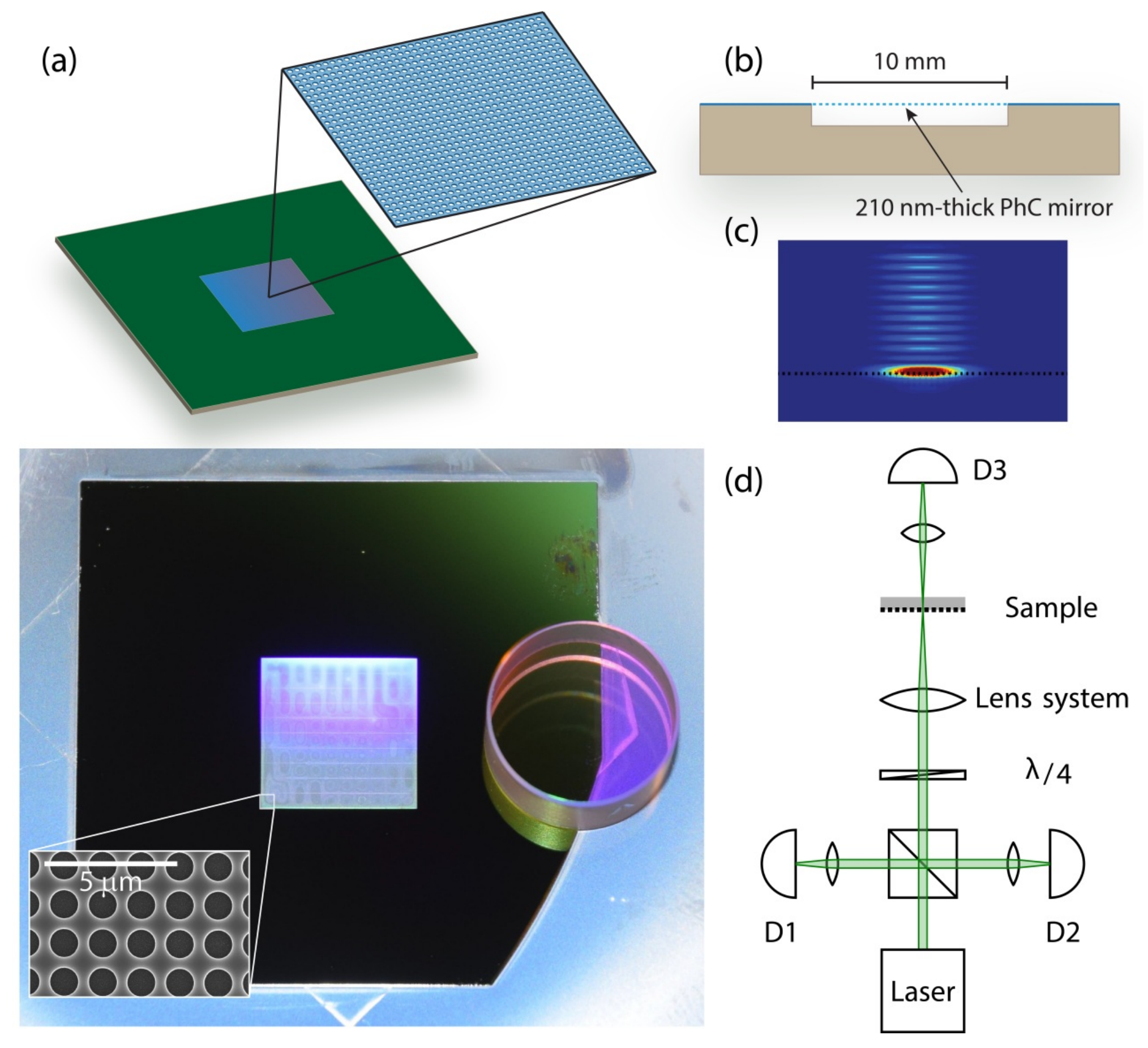}
	\caption{PhC mirrors and characterization setup. (a) Sketch of a suspended PhC mirror (top) and a photograph of a \SI{10}{mm}-wide, \SI{210}{nm}-thick PhC mirror next to a commercial 1/2 inch mirror for size comparison (bottom). The rectangular shaped patterns within the PhC are stitching errors from the mainfields of the beamwrite, which do not affect the measured reflectivity significantly. The inset shows a scanning electron microscope picture of the actual photonic crystal. The full mirror is made up of around \num{6e7} holes. (b) Illustration of the cross section of the mirror. The thin membrane is made of SiN and is supported by a silicon chip. (c) FDTD simulation of a reflected light mode on a PhC membrane. (d) Simplified setup used to characterize the reflection of PhC mirrors. We focus a wavelength-tunable laser beam perpendicularly onto the membrane. The radius of the incident beam is controlled with a lens system. For each radius, we acquire the reflection (D2) and transmission (D3) in relation to a reference beam (D1) that is split off the laser output using a polarization beam splitter. More details can be found in Appendix A.}
	\label{fig:setup}
\end{figure}

Our suspended PhC mirrors are fabricated from high-stress (\SI{1}{GPa}) LPCVD SiN films deposited on \SI{200}{\um} Si wafers. The geometry of the PhC structures is optimized for each desired film thickness to a wavelength of \SI{1550}{nm} using finite-difference time-domain (FDTD) simulations \cite{Fan2002}. The structures are patterned on the SiN films using electron beam lithography and a plasma etching process ($\mathrm{CHF_3+O_2}$). Stitching errors occur about every millimeter due to stage drifts during the beamwrite and are on the order of \SI{1}{\um} wide. One of the fundamental challenges for large aspect-ratio membranes is suspending them without causing any fractures. Typically liquid etchants such as KOH or TMAH are used to release free-standing SiN structures from their Si substrates. However, such wet processes produce a number of forces, like turbulences and surface-tension at the interfaces, which can easily destroy the fragile suspended PhC mirrors. These methods also leave residues that negatively impact the optical performance of the mirrors, requiring additional liquid cleaning steps that decrease the the fabrication yield, specially for large area devices~\cite{Norte2014}. In order to overcome these limitations, we have developed a stiction-free RIE-ICP plasma release using $\mathrm{SF_6}$~\cite{Patent2017}. A hot piranha solution consisting of sulfuric acid and hydrogen peroxide is first used to remove surface contaminations on the unreleased structures. This is followed by a diluted HF solution to smoothen the SiN surface~\cite{Ji2017} and remove surface-oxide from the silicon which allows for an even release of the membrane. Finally, we suspend the PhC mirrors using the $\mathrm{SF_6}$ plasma release. Figures~\ref{fig:setup}(a)-(c) show a sketch and a photograph of a \SI{10}{mm}-wide, \SI{210}{nm}-thin mirror, a scanning electron microscope picture of the PhC pattern, and numerical simulations of the reflected optical field. 

In order to characterize the optical properties of the PhCs, we fabricate three devices with different thickness and size:\ two \SI{210}{nm}-thick mirrors, \SI{4 x 4}{mm} and \SI{10 x 10}{mm}-large; and a \SI{56}{nm}-thick, \SI{1.6 x 1.6}{mm}-large one. We measure the mechanical spectrum of the fabricated devices to match those of bare square membranes with the same intrinsic stress. This indicates that despite the unconventionally large areas, the material stress remains high, which should guarantee the membrane's flatness, an important point when developing high reflectivity mirrors. This is also a relevant observation when estimating the thermal displacement noise of this type of device. More details can be found in Appendix B.

The devices are characterized by focusing a wavelength-tunable laser beam perpendicular to the PhC mirrors (\fig~\ref{fig:setup}(d)). We measure the reflected and transmitted power and compare it to a reference beam that does not interact with the devices. For calibration, we use a commercial broadband mirror at the same position as the PhC mirrors. The laser is tuned from \num{1530} to \SI{1630}{nm}. The recorded signal is normalized to the reference arm and to the calibration mirror to obtain the reflectivity spectra for multiple beam waists. We vary the beam radius between \SI{8}{\um} and \SI{1.1}{mm} using a lens system placed in front of the PhC membranes. This allows us to analyze the behavior of PhC membranes with different thicknesses to laser beams of varying sizes.

\begin{figure}[t]
	\centering
	\includegraphics[width = 1\columnwidth]{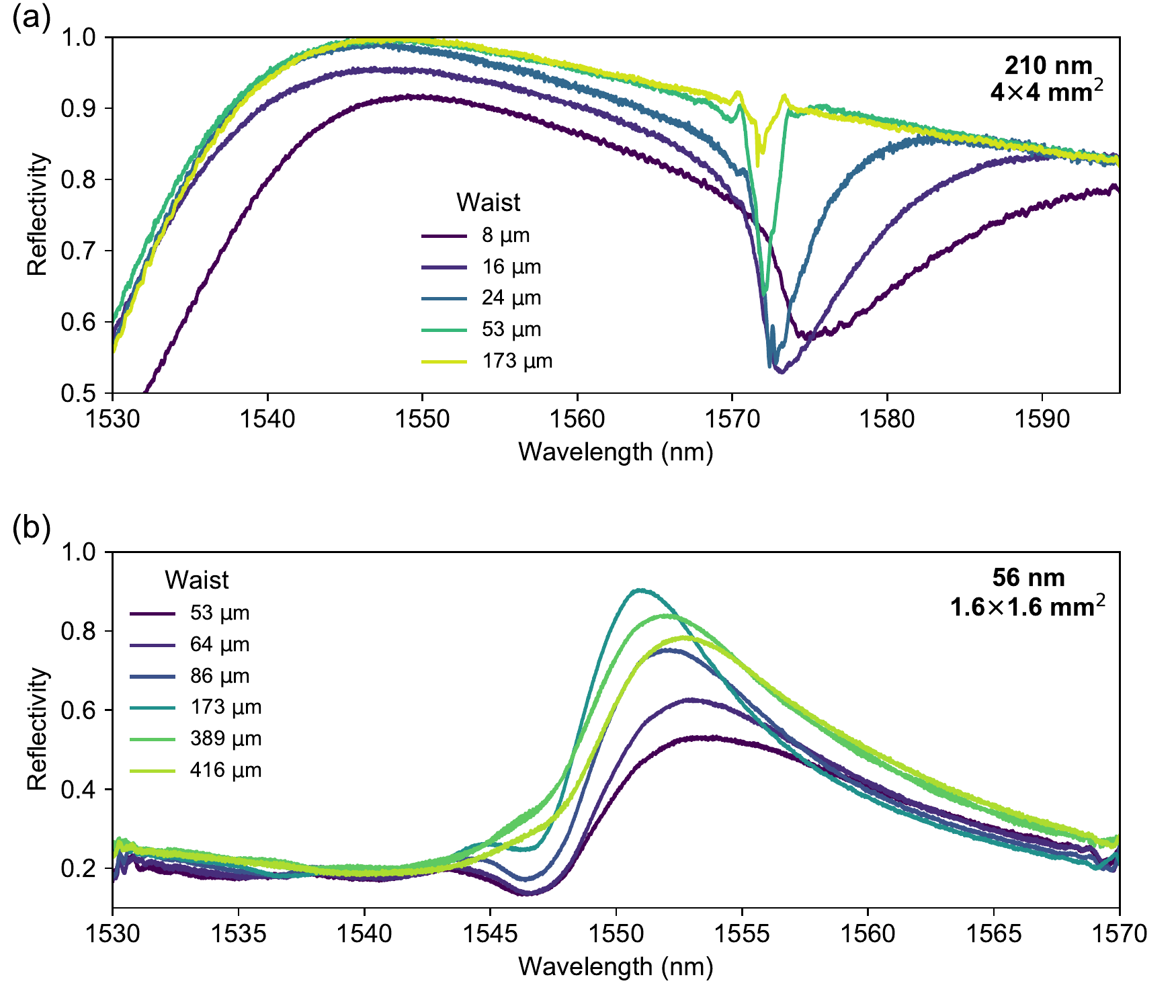}
	\caption{Reflectivity spectra of the PhC mirrors. Shown is a selection of measured reflectivity spectra of PhC mirrors with film thickness of (a) \SI{210}{nm} and (b) \SI{56}{nm}. Each spectrum shows the reflection of a Gaussian beam with the specified waist. As the waist increases, the incident beam approaches the behavior of a plane wave, for which the devices are optimal, and so the maximum reflectivity increases. Due to the finite size of the \SI{56}{nm} PhC mirror, its reflectivity drops as the incident beam becomes larger than the PhC area. The data were digitally processed to remove parasitic interferences from the substrate (see Appendix C for details).}
	\label{fig:spectra}
\end{figure}

Figure~\ref{fig:spectra} shows a selection of measured spectra of the \SI{210}{nm}-thick, \SI{4 x 4}{mm}-large and the \SI{56}{nm}-thick devices. While the PhC is completely released, for testing purposes the chip itself is not fully etched through (cf.\ \fig~\ref{fig:setup}(b)). This results in a parasitic interference pattern with a periodicity of \SI{1.8}{nm} on top of the expected PhC spectra, corresponding to a \SI{200}{\um}-thick Si etalon. Since the frequency of this interference is well defined, we post-process it by band-pass filtering the data. Appendix C contains the full set of acquired spectra, as well as a detailed description of the data processing.

The spectra of the \SI{210}{nm}-thick PhC mirror, shown in \fig~\ref{fig:spectra}(a), exhibit a resonance at \SI{1549}{nm} that varies little with the incident beam waist. At \SI{1573}{nm} a parasitic resonance emerges whose width increases as the waist becomes smaller. This can be understood by considering the decomposition of a Gaussian beam with waist $w_0$ into plane waves~\cite{Born1986}. The decomposition in terms of incidence angle is weighted by a Gaussian distribution with a standard deviation equal to the beam divergence $\theta = \lambda/\pi w_0$. A large waist $w_0$ has a small divergence $\theta$, which is a good approximation to a plane wave with a zero angle of incidence. As $w_0$ decreases, $\theta$ becomes larger, and so plane waves with larger angles of incidence have a stronger weight on the decomposition. These waves can couple to PhC modes other than the resonance of interest, giving rise to parasitic features such as the one observed. We can apply the same reasoning to explain the increase in maximum reflectivity of the main resonance:\ the device geometry was optimized assuming a plane wave with normal incidence. Hence, beams with a large waist approximate this condition better, which results in a reflectivity closer to the optimized one.

In \fig~\ref{fig:spectra}(b) we observe that the main resonance of the \SI{56}{nm}-thick membrane exhibits stronger shifts in wavelength, width and maximum reflectivity with varying beam waist, in comparison to the \SI{210}{nm}-thick device. As explained in the work of Bernard et al.~\cite{Bernard2016}, the spectral response of plane waves incident on a PhC mirror depends on the angle of incidence. This dependence is stronger for thinner devices and results in large resonance wavelength shifts. Therefore, as the beam waist decreases -- and its divergence increases -- the reflectivity of thin devices is more strongly attenuated. On the other hand, due to the small size of the \SI{56}{nm}-thick PhC mirror (area of \SI{1.6 x 1.6}{mm} vs.\ \SI{4 x 4}{mm} for the \SI{210}{nm}-thick device)  beams with waists larger than \SI{280}{\um} are partially scattered outside the PhC and exhibit a decreasing maximum reflectivity (see below).

\begin{figure}
	\centering
	\includegraphics[width = 1\columnwidth]{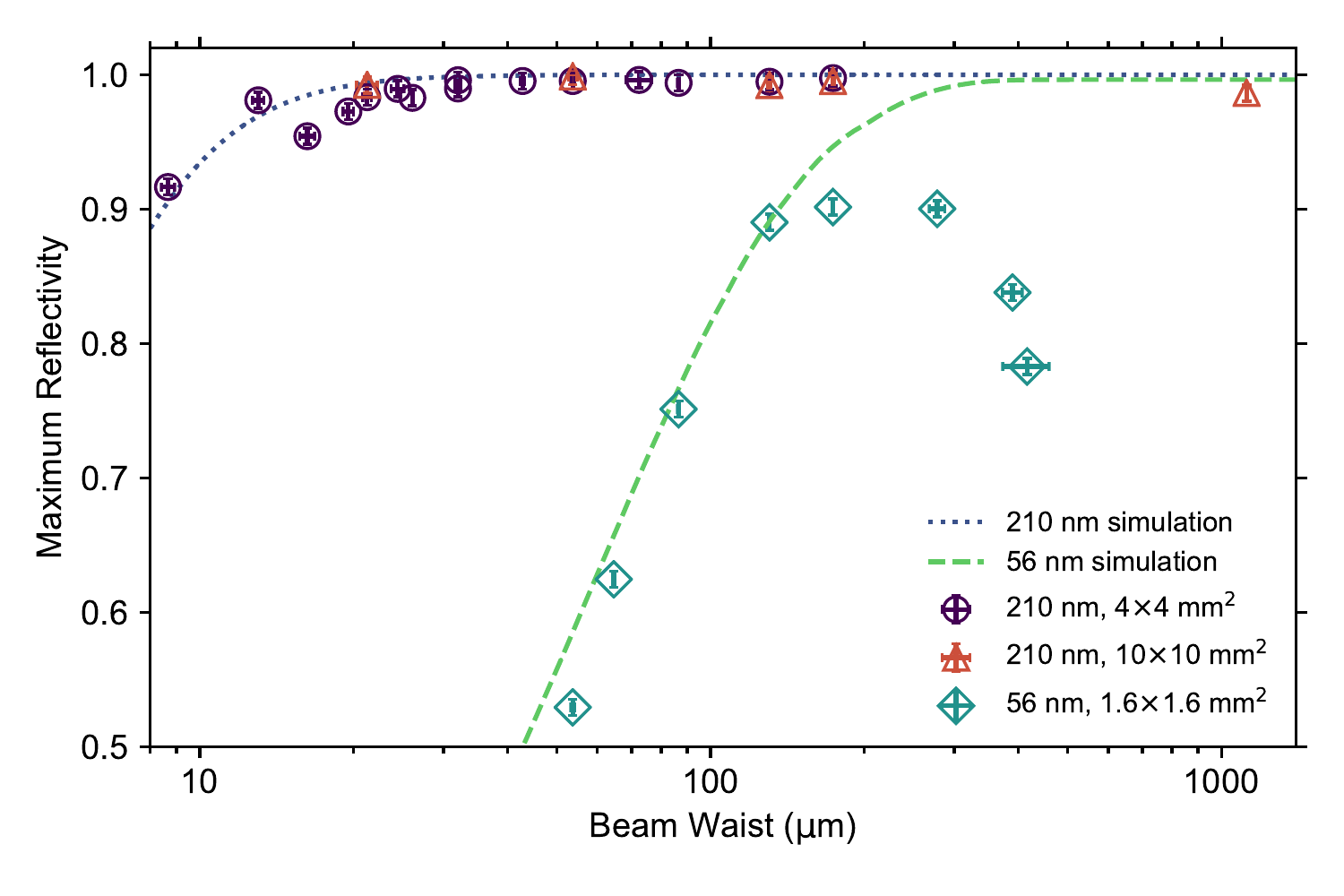}
	\caption{Maximum reflectivity as a function of incident beam waist. We show the maximum reflectivity for several PhC thicknesses and membrane sizes, highlighting the potential of these structures as large-area, high-reflectivity mirrors. For comparison, we include simulations, represented by lines, obtained from plane wave decomposition of Gaussian beams using RCWA. The reflectivity of the \SI{56}{nm}-thick PhC mirror decreases when the optical beam becomes comparable to the PhC diameter, underlining the importance of finite size effects. The measured data have an uncertainty in reflectivity of \SI{\pm 0.6}{\percent}. The uncertainty in the beam radii results from propagating the estimated uncertainties in the positions of the lens sets. For more details see Appendix A.}
	\label{fig:max_ref}
\end{figure}

Figure~\ref{fig:max_ref} shows the maximum reflectivity of all PhC membranes as a function of the incident beam waist. As described in the previous paragraphs, larger beam waists approximate the design conditions of the PhCs better, reducing the amount of light that couples to unwanted modes. As such, the maximum reflectivity increases for larger beams. We verify this behavior with simulations and plane wave decomposition:\ starting with the geometry parameters that resulted from the FDTD optimization and that were patterned on the SiN membranes, we simulate the reflectivity of plane waves with varying angles of incidence at the resonance wavelength using rigorous coupled-wave analysis (RCWA) \cite{Moharam1981}. The reflectivity at each beam waist is then the weighted sum of the simulation results, following a Gaussian distribution with standard deviation $\theta$, which is further described in Appendix D.

The reflectivity of the \SI{56}{nm}-thick PhC mirror decreases when the beam radius measures between \num{280} and \SI{390}{\um}. Considering that the field amplitude of a Gaussian beam falls as $e^{-r^2/w_0^2}$, where $r$ is the distance from the beam's center, we expect \SI{99}{\percent} of the field to be within a diameter of $6\times w_0$. Since the PhC measures \SI{1.6 x 1.6}{mm}, beams with waists larger than $1.6 / 6\,\si{mm} = \SI{270}{\um}$ will have larger field components reflecting off the area outside the PhC. This allows us to observe a smooth transition between two regimes:\ one, for small waists, where the PhC response is limited by a large beam divergence and another one for large waists, where the limitation is the finite size of the device. Between these two bounds we see a plateau where the maximum reflectivity \SI{>90}{\percent} is approximately constant. To the best of our knowledge, this is the highest reported reflectivity of a \SI{56}{nm} suspended PhC mirror, operating in a regime with lower beam divergence and finite size limits.

In conclusion, we fabricate and characterize the first suspended PhC mirrors that span areas up to square-centimeters, with reflectivities exceeding \SI{99}{\percent}, which are only limited by our measurement precision. Previous attempts focused on devices not wider than \SI{300}{\micro m}, resulting in strong limits to the maximum achievable reflectivity, in particular for devices thinner than $0.13\lambda$. By measuring the reflectivity spectrum of the PhC mirrors for varying incident beam waists, our work shows that these devices are indeed strongly affected by finite size effects. In particular, we observe a reflectivity of \SI{90}{\percent} for a $0.036\lambda$-thick PhC mirror at a wavelength of \SI{1550}{nm}, whereas previous reports of devices with similar thickness were limited to \SI{62}{\percent}~\cite{Bui2012,Bernard2016}. Despite the presence of fabrication errors in the lithography process for the largest device (see \fig~\ref{fig:setup}(a)), we suspect that its performance becomes insensitive to small defects in the PhC lattice for large beam waists, since the reflectivity shows no appreciable changes with increasing waists (see \fig~\ref{fig:max_ref}, 210~nm data). Since larger incident beams sample a larger area of the PhC structure, we observe that the reflectivity seems robust to imperfections in the 2D array of holes that arise from drifts or poor stitching during lithography. It is also important to note that these mirrors could be improved even further with more sophisticated methods of lithography. Electron beam lithography is prone to stage drifts during exposure and secondary back-scattering of electrons from the substrate which produce uneven dosing -- both of these effects can lead to an inhomogeneous lattice constant and variations in hole size. We envision scaling these devices further to full wafer sizes by using techniques such as nano-imprint lithography, lithography stepping, or interference lithography~\cite{Lu2010a}.

Due to its high intrinsic stress, LPCVD SiN PhC membranes should remain relatively flat even at larger areas. Together with the low optical absorption of SiN, this is a promising platform for light sails in future space probes propelled by light, such as the Breakthrough Initiative Starshot~\cite{Starshot2017}. In addition, high-stress SiN membranes have been shown to have high thermal noise suppression which becomes better with thinner, larger membranes~\cite{Norte2016}. At large scales, SiN PhCs could thus be an interesting route towards low-noise suspended mirror coatings. In Appendix E, we include a first-order estimate of the thermal displacement noise such devices would have, and make a baseline comparison of their performance to the well-documented mirror coatings used on LIGO test-masses.

Furthermore, the fact that these mirrors are suspended allows them to be used in a variety of applications that profit from mechanical tuning of mirrors. Deformable mirrors could be realized with these PhC structures~\cite{Madec2012}, for example through electrostatic-tuning with arrays of electrodes close to the mirror, or even as displacement noise tunable mirrors, using techniques such as optomechanical feedback control \cite{Wiseman1994}. Further experiments are planned to study the transversal mode composition of the reflected beam and to properly characterize the devices' optical absorption and scattering using a high finesse optical cavity. These developments open up a new paradigm in photonics -- one that steers away from the focus on simply miniaturizing components, but instead tries to bring the performance of nano-engineered materials to large scales.

\begin{acknowledgments}
We thank F.\ Alpeggiani, P.\ Klupar, A.\ Loeb, and P.\ Worden for helpful discussions. We also acknowledge valuable support from the Kavli Nanolab Delft, in particular from C.\ de Boer and M.\ Zuiddam. This project was supported by the European Research Council (ERC StG Strong-Q) and by the Netherlands Organisation for Scientific Research (NWO/OCW), as part of the Frontiers of Nanoscience program.
\end{acknowledgments}

\clearpage

\setcounter{figure}{0}
\renewcommand{\thefigure}{A\arabic{figure}}
\setcounter{equation}{0}
\renewcommand{\theequation}{A\arabic{equation}}

\section*{Appendix A: Setup}

\begin{figure}
	\centering
	\includegraphics[width = 1\columnwidth]{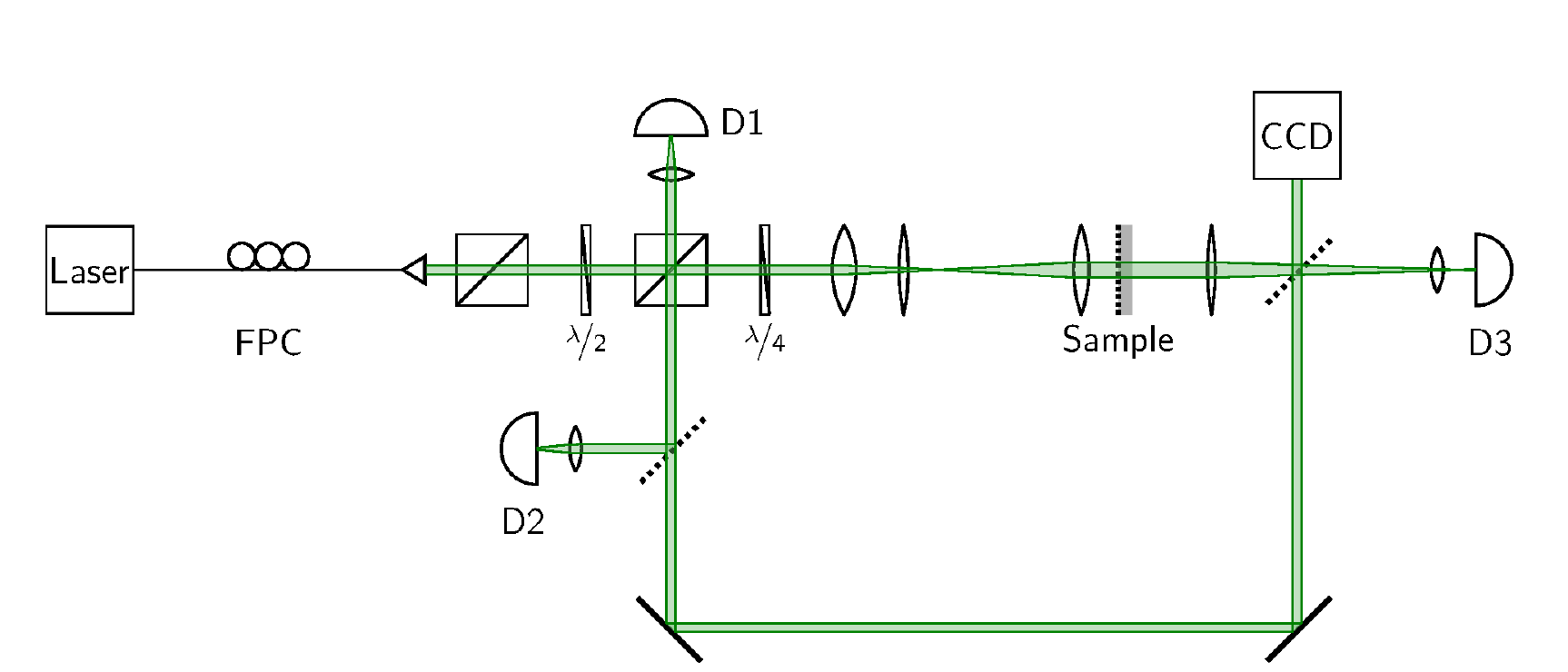}
	\caption{Schematic of the complete characterization setup. See text for a detailed description.}
	\label{fig:sup_setup}
\end{figure}

In order to characterize the optical properties of the PhCs, we use the setup outlined in \fig~\ref{fig:sup_setup}. A fiber coupled Santec TSL510 wavelength-tunable laser is used as a light source. Its polarization is adjusted with a fiber polarization controller (FPC) and the beam is brought into free-space using a triplet collimator (Thorlabs TC12APC-1550). The first polarization beam splitter (PBS) is used as a polarizer to mitigate the effect of polarization drifts in the fiber part of the setup. The second PBS splits light into the beam that is going to interact with the sample (incident beam), and a reference beam that is detected by D1.
We use the reference beam to remove the effect of power oscillations from the measured spectra that are not related to the sample. Between the two PBSs we place a half-waveplate ($\lambda/2$) to control the power ratio between the incident and reference beams. After the second PBS, the incident beam is focused onto the sample using a lens set (between 1 and 3 lenses) which are adjusted to change the waist from \num{8} to \SI{420}{\um}. Light reflected by the sample is sent into the second PBS. We place a quarter-waveplate ($\lambda/4$) in the incident beam path such that the reflected beam is separated from the incident beam by the PBS and then detected by D2. Light transmitted by the sample is recorded by D3.

For every lens system a commercial broadband-coated mirror is placed at the sample position and record reference and reflection spectra. These serve as calibration and remove effects such as losses in the optical path from the sample reflectivity spectra. The sample is then placed back and care is taken to ensure good tip/tilt alignment with respect to the optical beam, since the response of the PhC is very sensitive to the incident angle. Using flip mirrors we are also able to send the transmitted or reflected beam into an infra-red CCD. The camera helps during the tip/tilt alignment of the sample, or acts as a reference during the alignment of the lens system.

The reflectivity spectrum $R$ is calculated as
\begin{equation*}
	R = \frac{V_\text{PhC}/V_\text{PhC}^\text{ref}}{V_\text{cal}/V_\text{cal}^\text{ref}} \times R_\text{cal},
\end{equation*}
where $V_\text{PhC}$ and $V_\text{cal}$ are the voltage signals of the reflected beams from the PhC and the calibration mirror, $V_\text{PhC}^\text{ref}$ and $V_\text{cal}^\text{ref}$ are the reference signals of the corresponding PhC and calibration mirror measurements, and $R_\text{cal}$ is the reflectivity of the calibration mirror, which is specified to be \SI{99.8 \pm .3}{\percent}. We consider all measurement uncertainties to be independent from each other, and estimate the uncertainty in reflectivity $\Delta R$ via the method of uncertainty propagation
\begin{equation*}
	\Delta R = \sqrt{\sum_x \robra*{\frac{\partial R(x)}{\partial x} \Delta x}^2} \approx 0.006,
\end{equation*}
where $x = \{ V_\text{PhC}, V_\text{PhC}^\text{ref}, V_\text{cal}, V_\text{cal}^\text{ref}, R_\text{cal} \}$.

To estimate the uncertainty of the beam waist, the same propagation method was applied.
As uncertainty parameters, the position of each lens (with a count varying between 1 and 3) and its focal length is taken into account. The beam was propagated through the lens system by means of the complex beam parameter and ABCD matrices.

The photodetectors D1, D2, D3 are home-built surface-mount-device circuits equipped with a JDSU ETX500 photodiode. By means of electronic design and spectral characterization, a linear response to the optical input power is guaranteed.

\begin{figure*}
	\centering
	\includegraphics[width=0.8\textwidth]{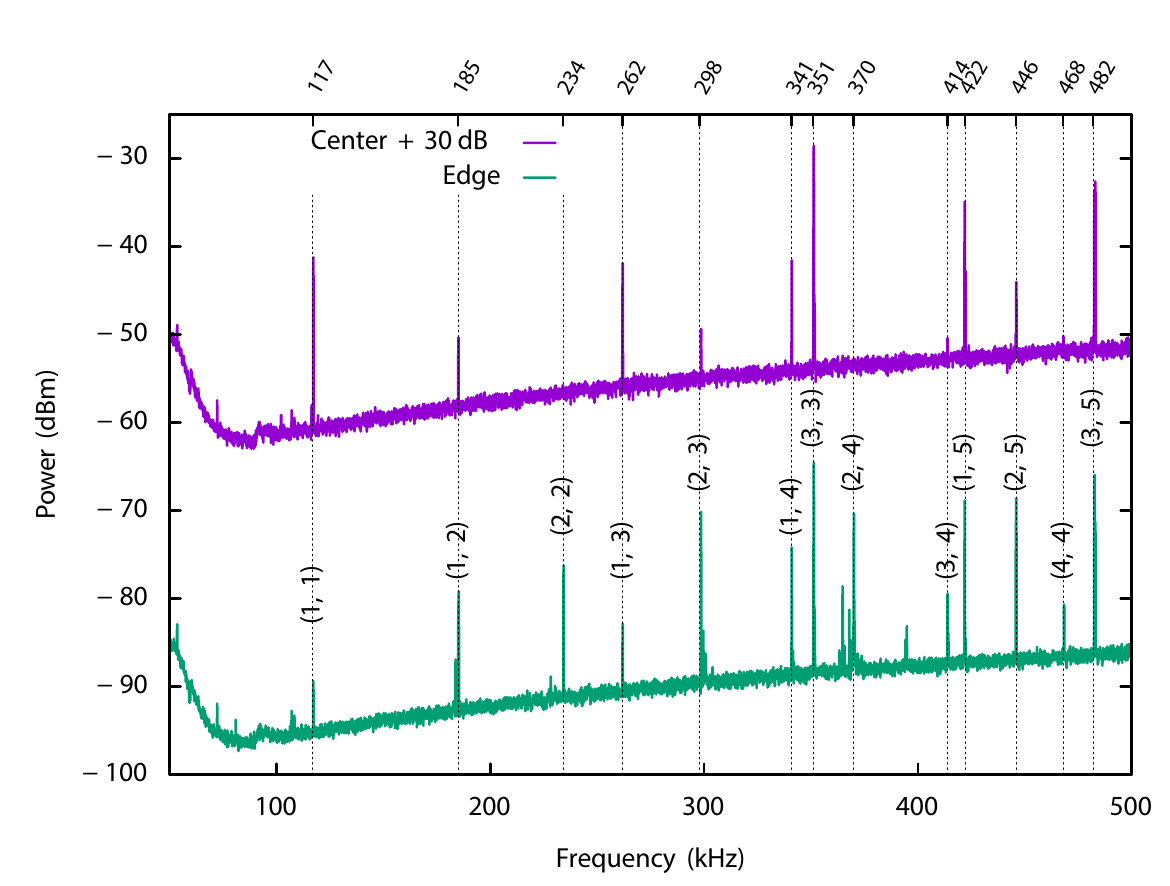}
	\caption{Measured mode spectrum. Shown is the spectrum of the back reflected light from a \SI{4 x 4}{\mm} PhC sample detected with a homodyne detector. The optical power for this measurement was set to \SI{1}{\mW}, the resolution bandwidth to \SI{100}{\Hz} and the spectrum was averaged 50 times. The device was driven with a piezo actuator connected to a white noise generator (peak-to-peak voltage \SI{100}{\mV}) inside a vacuum chamber at \SI{1e-5}{\milli\bar}. Electronic and
		displacement noise from the mounting-frame were subtracted from the displayed data.	We calibrated the noise from the mounting-frame by measuring a spectrum with the laser beam focused on the frame. The purple trace (top) shows data obtained with the laser focused in the center of the device, while for the green trace (bottom) the laser was focused onto the edge of the PhC mirror. The latter was done to record modes which have no net-effect on the reflected beam, i.e.\ a (2, 2) mode. The theoretically expected mode frequencies are highlighted on the upper horizontal axis and match the measured spectrum very closely.}
	\label{fig:sup_mech_spec}
\end{figure*}

\begin{figure*}
	\centering
	\includegraphics[width = 1\textwidth]{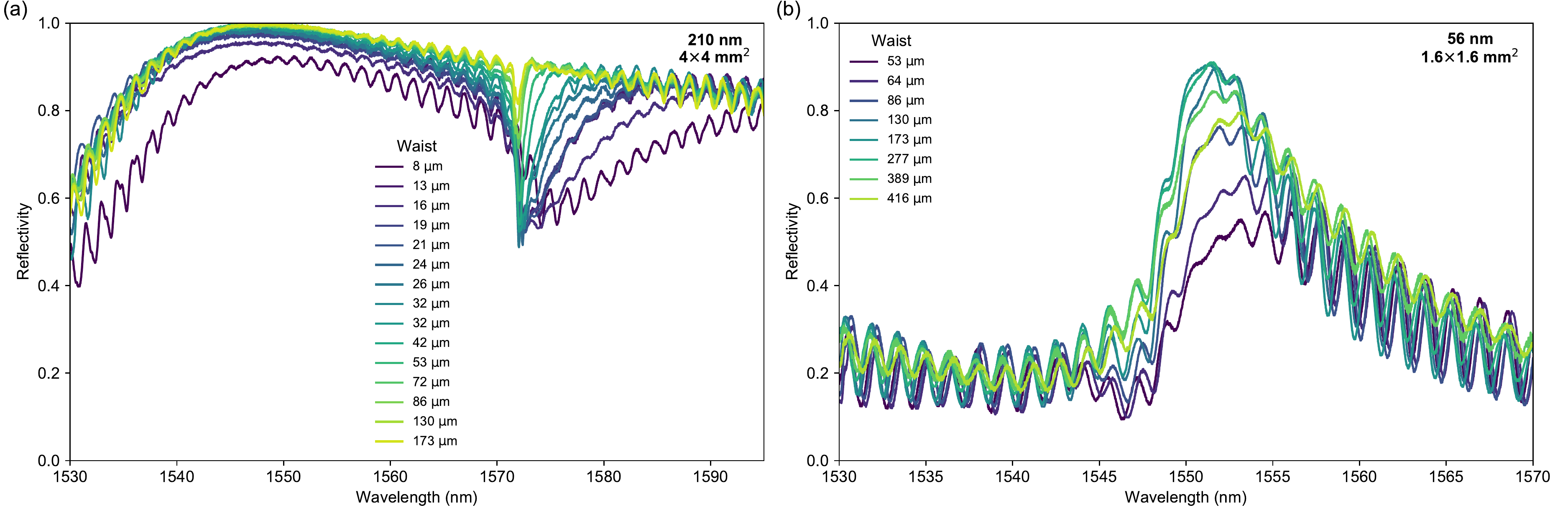}
	\caption{Unprocessed reflectivity spectra of the PhC mirrors.}
	\label{fig:sup_unprocessed_spectra}
\end{figure*}

\begin{figure*}
	\centering
	\includegraphics[width = 1\textwidth]{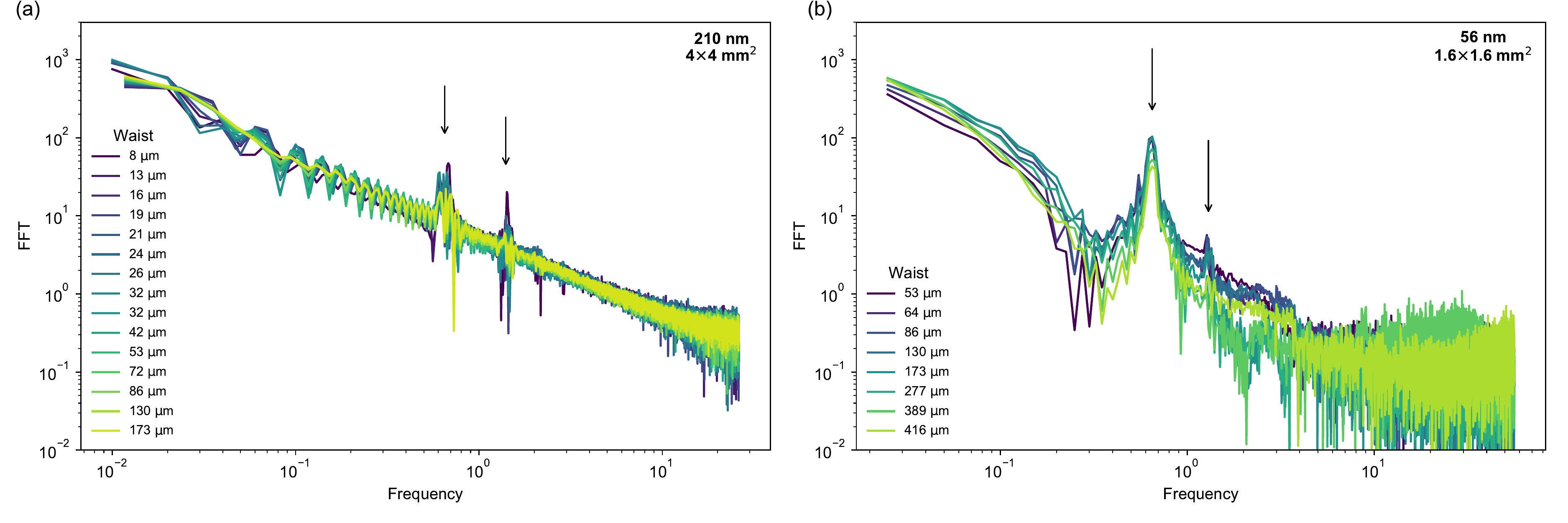}
	\caption{Fourier transformations of the unprocessed optical spectra.}
	\label{fig:sup_fft}
\end{figure*}

\begin{figure*}
	\centering
	\includegraphics[width = 1\textwidth]{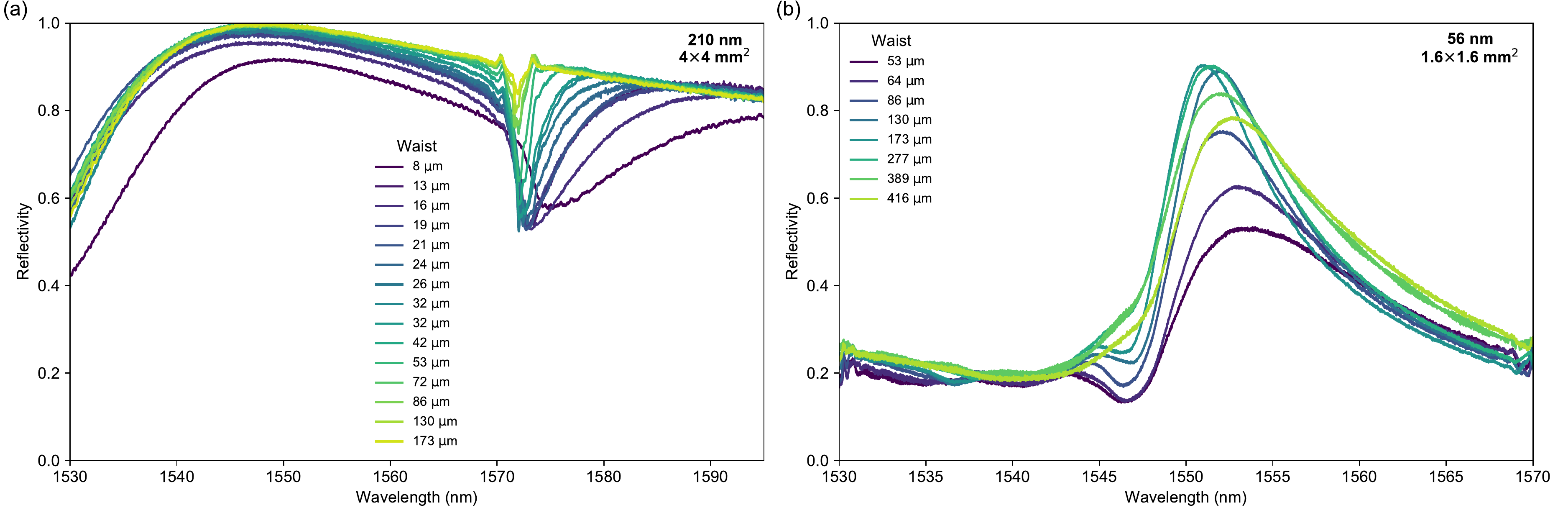}
	\caption{Filtered reflectivity spectra of the PhC mirrors.}
	\label{fig:sup_filtered}
\end{figure*}

\section*{Appendix B: Mechanical modes of PhC mirror}

In order to get a better understanding of the mechanical properties of the PhC mirrors, a displacement spectrum was recorded for a \SI{4 x 4}{\mm} mirror. The device was mounted in a vacuum chamber and a homodyne detector measured the phase quadrature of the back reflected light. Figure~\ref{fig:sup_mech_spec} shows a spectrum recorded in the center of the mirror and another spectrum measured at the edge of it. The off-center spectrum displays mechanical modes which have no net-effect on centered beams, i.e.\ modes which are mirror symmetric on two orthogonal axes. The fundamental mechanical frequency of a bare square membrane is given by
\begin{equation}
\omega_\text{m} = \frac{2 \pi}{\sqrt 2 L} \sqrt{\frac{\sigma}{\rho}},
\label{eq:freq_square_membrane}
\end{equation}
where $L$ is the length the membrane, $\rho$ is the material density (\SI{2.7}{g/cm^3} for SiN), and $\sigma$ is the tensile pre-stress in the film, which we take to be \SI{1}{GPa}, a value defined by the parameters of the LPCVD deposition process. A square membrane with the same dimensions and material properties as the one discussed here has a fundamental frequency of \SI{108}{kHz}. Comparing this to the value measured for our device (\SI{117}{kHz}) leads us to conclude that the pre-stress remains high even for these unconventionaly large suspended areas.

To simplify the analysis, we model the structure as a simple square membrane without holes. 
Starting from the fundamental mode frequency of the device, we can calculate the frequencies of all higher-order modes~\cite{Chakram2014}, which we plot on the upper horizontal axis in \fig~\ref{fig:sup_mech_spec}. We observe a very accurate fit of the measured higher-order modes (the deviation is around \SI{1}{\percent}). The measured spectrum also allows to predict the noise performance of an optical cavity made with such PhC mirrors~\cite{Aspelmeyer2014}.\\

\section*{Appendix C: Post-processing of spectral data}

Figure~\ref{fig:sup_unprocessed_spectra} shows the full, unprocessed set of measured reflectivity spectra for the (a) \SI{210}{nm} and (b) \SI{56}{nm}-thick devices. We measured the spectra for several incident beam waists, which are indicated in the figure legends. The spectra follow the expected Fano resonance shape, characteristic for this type of device.

In addition, we also see a parasitic oscillation with a periodicity of \SI{1.8}{nm}. This is because the devices are suspended but the substrate is not etched through. In fact, to facilitate the testing process, the membranes are undercut by only a few \si{\um} on top of the \SI{200}{\um} silicon substrate (cf.\ \fig~\ref{fig:setup}(b)). The observed periodic pattern arises from interference of reflections from the substrate interfaces. Using the thickness of the silicon substrate and a refractive index of \num{3.5}, we calculate a free spectral range of \SI{1.7}{nm} for a wavelength of \SI{1550}{nm}, which is in excellent agreement with the observed oscillations. The periodicity observed on the measured spectra is equal for all measurements and as such, we remove it digitally using the procedure described below.

The Fourier transformations of the unprocessed reflectivity spectra show peaks (marked by the arrows in \fig~\ref{fig:sup_fft}) that are well defined and common to all measurements, which we associate with the described etalon effect. These are obtained with the \texttt{fft.rfft} and \texttt{fft.rfftfreq} commands of the Python \texttt{numpy} library. To remove these parasitic features from the Fourier transforms, we apply a Tukey filter, with filter parameter \num{0.9}, around the identified peaks. The filter was generated using the function \texttt{signal.tukey} from the \texttt{scipy} library. Finally, the filtered spectra are obtained by performing the inverse Fourier transform using the \texttt{fft.irfft} command. These can be seen in \fig~\ref{fig:sup_filtered}. We carefully verify that filtering does not change the reading of the maximum reflectivity.

\begin{figure*}
	\centering
	\includegraphics[width = 0.5\textwidth]{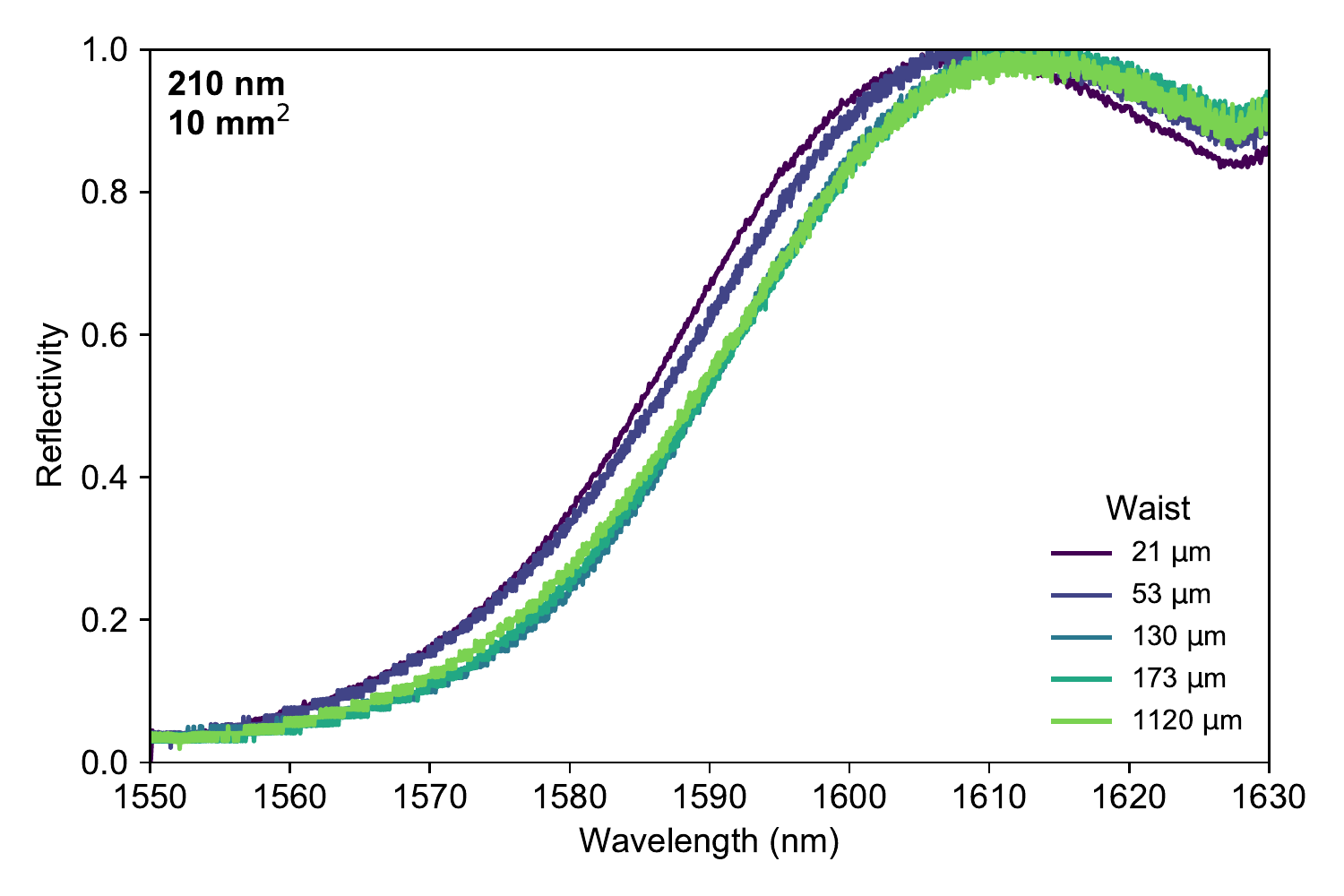}
	\caption{Unprocessed reflectivity spectra of the \SI{210}{nm}-thick, \SI{10}{mm}-wide device.}
	\label{fig:sup_unprocessed_10mm}
\end{figure*}

Figure~\ref{fig:sup_unprocessed_10mm} shows the raw data of the \SI{210}{nm}-thick, \SI{10}{mm}-wide device. This device was attached to a single-side polished carrier wafer using an index matching oil. Since the bottom surface of the substrate is rough, the interference between interfaces is no longer visible. Therefore this data set did not require any post-processing.

\section*{Appendix D: Simulated reflectivity spectra for Gaussian beams}

To simulate the expected reflectivity spectra of the PhC membranes, we use a rigorous coupled wave analysis (RCWA)~\cite{Moharam1981} combined with a plane wave decomposition~\cite{Rhodes1964}. On the one hand, choosing a RCWA implies that the simulated spectra are valid only for periodic structures, i.e.\ spectral changes caused by diffraction effects at the membrane's edge cannot be recreated. As this approach commonly starts with a plane wave as the incident electromagnetic field, it also requires to implement a composition of plane wave spectra when dealing with Gaussian beams. On the other hand, a finite element analysis (FEA) can in principle be set up to compute spectra of a finite, e.g.\ \SI{10 x 10}{\mm} large, PhC membrane excited by a Gaussian beam of waist $w_0$. However, the simplicity of a FEA approach has disadvantages when it comes to hardware, especially memory, requirements. To faithfully simulate the structure, the model volume has to cover about \SI{20 x 20 x 1}{\mm} and still capture the details of the nano-scale membrane, which increases the memory usage drastically. A RCWA only discretizes the actual PhC membrane, such that the reflected field can be retrieved at any point above or below the structure. Finally, having simulated a set of plane waves via RCWA allows to assemble Gaussian beams of any size. For our requirements, the RCWA approach therefore is most appropriate.
\\

We simulate two different unit cells:
\begin{itemize}
	\item \SI{210}{nm} thick, lattice constant \SI{1.355}{\um}, hole radius \SI{0.5014}{\um}
	\item \SI{56}{nm} thick, lattice constant \SI{1.526}{\um}, hole radius \SI{0.6265}{\um}.
\end{itemize}
Each cell, composed of 140 modes, is excited at a certain wavelength with $66^2$ plane waves of various polar and azimuthal angles of incidence. To verify that the structure is polarization insensitive, the wavelength- and angle scan is conducted for both $s$- and $p$-polarization. The simulation is built around an open source RCWA package~\cite{Liu2012}, which rotates the $s$ and $p$ component of the incident field with respect to the incident angles. To align the polarization uniformly for all angles, the $s$- and $p$-electric field component $E$ are transformed as
\begin{subequations}
	\begin{align}
		E_s &\mapsto \cos(\theta) / \cos(\phi) \\
		E_p &\mapsto -\sin(\theta)
	\end{align}
\end{subequations}
and
\begin{subequations}
	\begin{align}
		E_s &\mapsto \sin(\theta) / \cos(\phi) \\
		E_p &\mapsto \cos(\theta)
	\end{align}
\end{subequations}
for a $s$- and $p$-polarized beam, respectively. Here the polar angle is denoted by $\phi$, while $\theta$ is the azimuthal angle. This set of transformations inverts the global rotation implemented in the software package.

\begin{figure*}[ht]
	\centering
	\includegraphics[width = 1\textwidth]{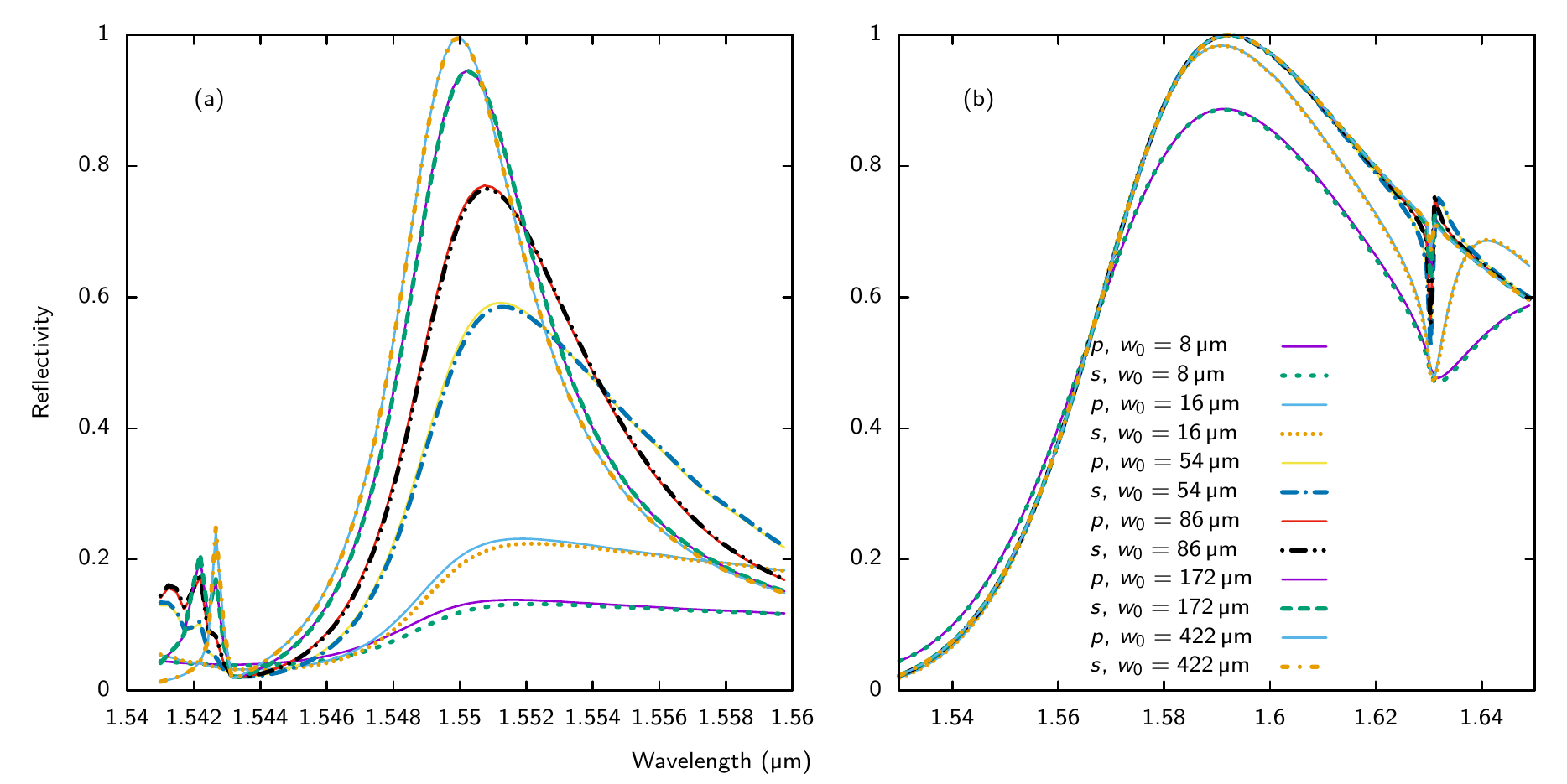}
	\caption{Reflectivity simulations for PhC membranes. Reflectivity spectra for $s$- and $p$-polarized Gaussian beams of waist $w_0$. Panel (a) and (b) show data for a \SI{56}{nm} and \SI{210}{nm} membrane, respectively.}
	\label{fig:rcwa_56210}
\end{figure*}

\begin{figure*}
	\centering
	\includegraphics[width = 1\textwidth]{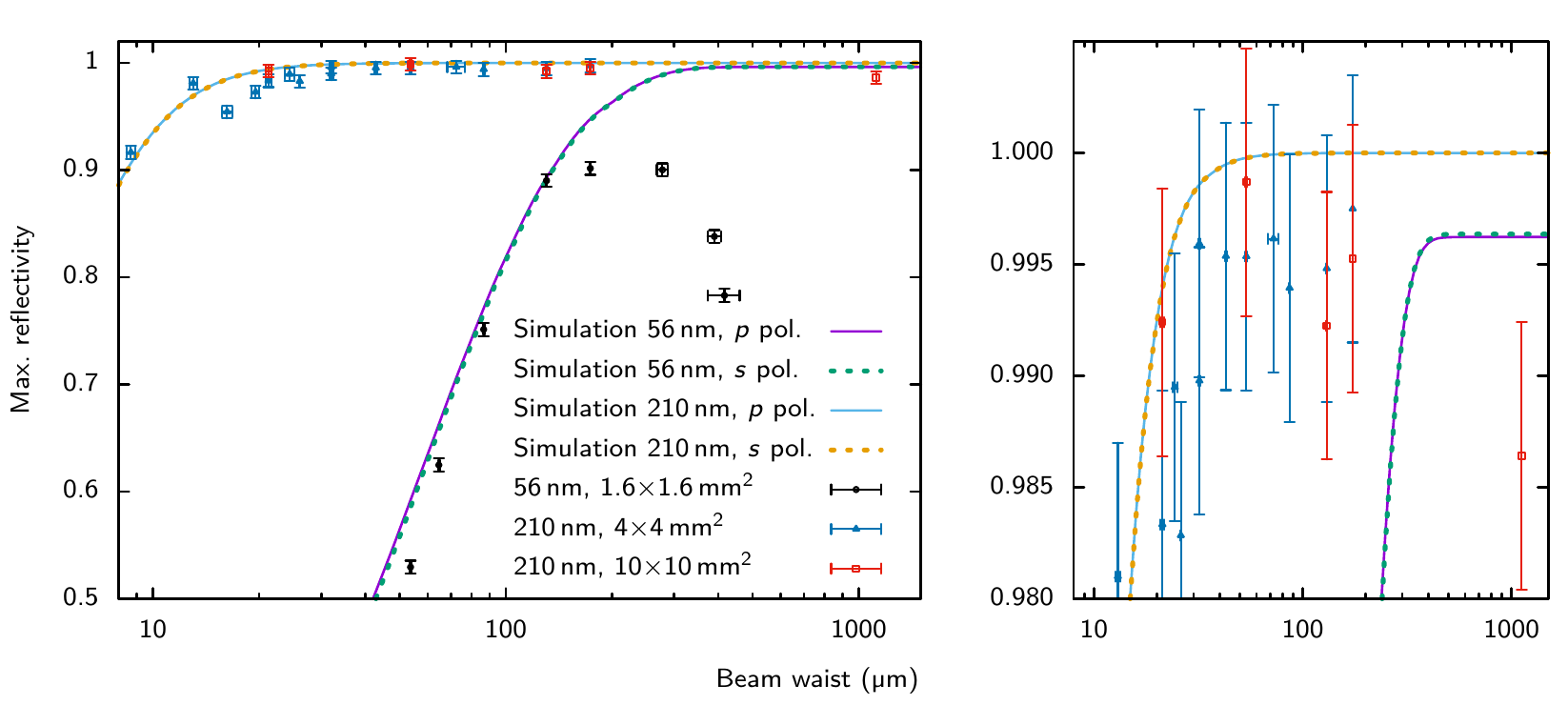}
	\caption{Extracted peak reflectivity compared to measured data. Shown is the same data as in the main text with additional simulations for various polarizations.}
	\label{fig:rcwa_rmax}
\end{figure*}

Having computed the reflection coefficients $r_{s, p}(\phi, \theta, \lambda) \in \mathbb C$ for a plane wave, the reflectivity of a Gaussian beam with the electric field distribution
\begin{equation}
E(x, y) = \sqrt{\tfrac{2}{\pi w_0^2}} e^{-\frac{x^2 + y^2}{w_0^2}}
\end{equation}
at the waist position ($z = 0$) is obtained by weighting the reflection coefficients according to the plane wave decomposition
\begin{equation}
E(k_x, k_y) = \iint_{-\infty}^{\infty} E(x, y) e^{\imath \robra{k_x x + k_y y - \frac{2 \pi c_0}{\lambda} t + \phi}} \mathrm dx \mathrm dy.
\end{equation}
Transforming the expression to spherical coordinates, the power reflectivity spectra are computed via
\begin{equation}
R_{s, p}(\lambda, w_0) = \frac{\iint \abs{r_{s,p}(\phi, \theta, \lambda)}^2 e^{-\frac{1}{2} \robra{w_0 r \sin(\phi)}^2} \sin(\phi) \mathrm d\theta \mathrm d\phi}
{\iint e^{-\frac{1}{2} \robra{w_0 r \sin(\phi)}^2} \sin(\phi) \mathrm d\theta \mathrm d\phi}.
\end{equation}
Figure~\ref{fig:rcwa_56210} illustrates the results of the simulations for a \SI{56}{nm} and \SI{210}{nm} unit cell with the parameters given above. Due to a small mismatch in the cell parameters the absolute resonance frequency of the measured and simulated spectra deviate from each other slightly, however the overall features of the measurement are recreated faithfully. The individual simulations of $s$- and $p$-polarization exhibit virtually no difference, which is important as we probe the membranes with a circularly polarized beam. From the data plotted in \fig~\ref{fig:rcwa_56210}, the maximum achievable reflectivity can be readily extracted by determining the peak reflectivity for each wavelength and beam waist scan. In figure \ref{fig:rcwa_rmax} the results from the main text are plotted again and the simulated maximum reflectivity is shown for both polarizations. A zoom into the last two percent helps reading out the measured values.

\section*{Appendix E: Estimation of thermal displacement noise}

Thermal mirror coating noise is currently one of the main limitations to the sensitivity of high-precision experiments including atomic clocks, frequency stabilized lasers and gravitational wave detectors. Widely used in these communities are distributed Bragg reflector (DBR) coatings made of alternating layers of dielectric materials. One challenge with DBR coatings is that their reflectivity has a natural trade-off with the number of layers used (and therefore the  total thickness). Increases in coating thickness are usually associated with increases in coating's thermal Brownian noise. Attaining high reflectivity requires DBR surfaces which are commonly microns thick, while PhC mirrors can realize similarly high reflectivities with a membrane which is only the thickness of a single DBR layer. Additionally, their suspended geometry makes them into an easily deformable mirror surface, which potentially provides mechanical isolation from substrate (i.e.\ test mass) noise, and the possibility to dynamically tune the mechanical properties using optomechanical techniques.

In order to assess the potential that suspended PhC membranes could have in these types of applications, we perform a simplified calculation of the thermal noise performance of suspended PhC mirrors in this section and compare it to the thermal coating noise of the a-LIGO experiment~\cite{LIGO2015}.

For large LPCVD SiN mirrors we estimate the biggest noise contribution to be due to the thermal displacement noise, which for a square membrane is given by~\cite{Saulson1990}
\begin{equation}
S_x(\omega) = 
\frac{4 k_B T m_\text{eff} \omega_\text{m} / Q}{m_\text{eff}^2 \robra{\robra{\omega^2 - \omega_\text{m}^2}^2 + \robra{\omega_\text{m} \omega / Q}^2}},
\end{equation}
where $Q$ is the mechanical quality factor, $k_B$ the Boltzmann constant, $T$ is the membrane's temperature, $m_\text{eff}$ the effective mass of the membrane's fundamental mode, and $\omega_\text{m}$ its frequency given by Eq.~\eqref{eq:freq_square_membrane}. The effective mass can be estimated with $m_\text{eff} = m / 4$~\cite{Wilson2012}, where $m$ is the physical mass. It is further reduced by a factor of \num{0.3}, which accounts for the mass lost to the PhC holes.

There are numerous limiting factors to the mechanical quality of a membrane~\cite{Schmid2011,Chakram2014}:\ thermoelastic damping, surface defects, Akhiezer damping, etc. For LPCVD SiN membranes, the most relevant factors are acoustic radiation losses and damping from collisions with gas particles. The acoustic radiation losses into the substrate generally scale as the ratio between  $L$ and the thickness $h$:\ $Q_\text{rl} \propto L / h$~\cite{Schmid2011,Chakram2014}. Considering previously measured radiation limited quality factors of \num{4e7} for $L/h = \num{5e4}$, we can extrapolate the radiation limit for arbitrary sizes of the PhC mirror. In addition, the quality factor limited by gas damping is given by~\cite{Bao2002}
\begin{equation}
Q_\text{p} = \left(\frac{\pi}{2}\right)^{\frac{3}{2}} \rho h \frac{\omega_\text{m}}{2\pi}\sqrt{ \frac{R T}{m_g} } \frac{1}{p},
\end{equation}
where $R$ is the ideal gas constant, $p$ is the pressure, and $m_g$ is the molecular mass of the background gas molecules. The final Q is given by $Q^{-1}=\sum_i Q_i^{-1}$, where $Q_i$ are the various contributions mentioned above.

In order to estimate the thermal displacement noise of suspended PhC mirrors, we need to assume some parameters. We chose those of the a-LIGO experiment, since its DBR coated mirrors are well known and characterized in terms of noise properties. This allows us to make a first-order comparison of how suspended PhC mirrors would operate when scaled up to the same size of the a-LIGO mirrors.

We take a SiN film thickness of \SI{210}{nm}, a lateral size $L = 350$~mm, and effective mass $m_\text{eff} = 12$~mg. We assume the mirrors are placed in a vacuum chamber at room temperature with a pressure of $p = 7\times10^{-9}$~hPa and that the main background gas component is hydrogen~\cite{LIGO2015}. These parameters result in a fundamental mechanical frequency of $\omega_\text{m}/2\pi = 1200$~Hz and a pressure-limited fundamental mode mechanical quality factor $Q_p \sim 10^8$. Considering a detection frequency of $100$~Hz, we find a thermal displacement noise of around $10^{-17}~\textrm{m} / \sqrt{\textrm{Hz}}$. For comparison, the thermal coating noise of the mirrors used in a-LIGO is of the order of $10^{-20}~\textrm{m} / \sqrt{\textrm{Hz}}$.

These estimates are mainly limited by the environment pressure, which sets an upper bound to the mechanical quality factor. By decreasing the pressure further it could be possible to improve the thermal displacement noise significantly. Furthermore, suspended mirrors have the additional advantage that they can be adjusted, for example either through the addition of tuning electrodes or optomechanical techniques, in order to further reduce the noise performance in the desired regime.

It is important to note that the focus of this calculation is the estimation of thermal Brownian noise associated with the mirror coatings. The presented calculations and mirror designs are heuristic in nature and only allow one to make estimates comparing different mirror coating noise performances. It is however not entirely clear for example, how the Brownian noise related to the substrate (or test mass) would couple to such a suspended mirror. This could be relevant for monolithic cavities in quantum optomechanics experiments at room temperature, where substrate thermal noise is the dominant source of heating in laser cooling experiments~\cite{Wilson2012}. In addition, it has been shown that at increasingly large aspect ratios the substrate thickness becomes a significant variable in a membrane's mechanical quality factor~\cite{Norte2016}. A massive substrate could work well in experiments which require large test masses. How these effects translate to the cm-scale remains an open question.


\begin{thebibliography}{10}
	\newcommand{\enquote}[1]{\textit{#1}}
	
	\bibitem{Fan2002}
	S.~Fan and J.~D. Joannopoulos, \enquote{Analysis of guided resonances in
		photonic crystal slabs,} Phys. Rev. B \textbf{65}, 235112 (2002).
	
	\bibitem{Kanskar1997}
	M.~Kanskar, P.~Paddon, V.~Pacradouni, R.~Morin, A.~Busch, J.~F. Young, S.~R.
	Johnson, J.~MacKenzie, and T.~Tiedje, \enquote{Observation of leaky slab
		modes in an air-bridged semiconductor waveguide with a two-dimensional
		photonic lattice,} Appl. Phys. Lett. \textbf{70}, 1438--1440 (1997).
	
	\bibitem{Lu2010}
	F.~Lu, F.~G. Sedgwick, V.~Karagodsky, C.~Chase, and C.~J. Chang-Hasnain,
	\enquote{Planar high-numerical-aperture low-loss focusing reflectors and
		lenses using subwavelength high contrast gratings,} Opt. Express \textbf{18},
	12606--12614 (2010).
	
	\bibitem{Fattal2010}
	D.~Fattal, J.~Li, Z.~Peng, M.~Fiorentino, and R.~G. Beausoleil, \enquote{Flat
		dielectric grating reflectors with focusing abilities,} Nat. Photon.
	\textbf{4}, 466--470 (2010).
	
	\bibitem{Guo2017}
	J.~Guo, R.~A. Norte, and S.~Gr\"{o}blacher, \enquote{Integrated optical force
		sensors using focusing photonic crystal arrays,} Opt. Express \textbf{25},
	9196--9203 (2017).
	
	\bibitem{Chen2017}
	X.~Chen, C.~Chardin, K.~Makles, C.~Ca\"{e}r, S.~Chua, R.~Braive,
	I.~Robert-Philip, T.~Briant, P.-F. Cohadon, A.~Heidmann, T.~Jacqmin, and
	S.~Del\'{e}glise, \enquote{High-finesse Fabry-Perot cavities with
		bidimensional Si$_3$N$_4$ photonic-crystal slabs,} Light Sci. Appl.
	\textbf{6}, e16190 (2017).
	
	\bibitem{Norte2016}
	R.~A. Norte, J.~P. Moura, and S.~Gr\"oblacher, \enquote{Mechanical resonators
		for quantum optomechanics experiments at room temperature,} Phys. Rev. Lett.
	\textbf{116}, 147202 (2016).
	
	\bibitem{Reinhardt2016}
	C.~Reinhardt, T.~M\"uller, A.~Bourassa, and J.~C. Sankey,
	\enquote{Ultralow-noise SiN trampoline MEMS for sensing and optomechanics,}
	Phys. Rev. X \textbf{6}, 021001 (2016).
	
	\bibitem{Bernard2016}
	S.~Bernard, C.~Reinhardt, V.~Dumont, Y.-A. Peter, and J.~C. Sankey,
	\enquote{Precision resonance tuning and design of sin photonic crystal
		reflectors,} Opt. Lett. \textbf{41}, 5624--5627 (2016).
	
	\bibitem{Kemiktarak2012}
	U.~Kemiktarak, M.~Durand, M.~Metcalfe, and J.~Lawall, \enquote{Cavity
		optomechanics with sub-wavelength grating mirrors,} New J. Phys. \textbf{14},
	125010 (2012).
	
	\bibitem{Bui2012}
	C.~H. Bui, J.~Zheng, S.~W. Hoch, L.~Y.~T. Lee, J.~G.~E. Harris, and C.~W. Wong,
	\enquote{High-reflectivity, high-{Q} micromechanical membranes via guided
		resonances for enhanced optomechanical coupling,} Appl. Phys. Lett.
	\textbf{100}, 021110 (2012).
	
	\bibitem{Campbell2017}
	S.~L. Campbell, R.~B. Hutson, G.~E. Marti, A.~Goban, N.~Darkwah~Oppong, R.~L.
	McNally, L.~Sonderhouse, J.~M. Robinson, W.~Zhang, B.~J. Bloom, and J.~Ye,
	\enquote{A Fermi-degenerate three-dimensional optical lattice clock,}
	Science \textbf{358}, 90--94 (2017).
	
	\bibitem{Kessler2012}
	T.~Kessler, C.~Hagemann, C.~Grebing, T.~Legero, U.~Sterr, F.~Riehle, M.~J.
	Martin, L.~Chen, and J.~Ye, \enquote{A sub-40-mHz-linewidth laser based on a
		silicon single-crystal optical cavity,} Nat. Photon. \textbf{6}, 687--692
	(2012).
	
	\bibitem{Reid2016}
	S.~Reid and I.~W. Martin, \enquote{Development of mirror coatings for
		gravitational wave detectors,} Coatings \textbf{6}, 61 (2016).
	
	\bibitem{Madec2012}
	P.-Y. Madec, \enquote{Overview of deformable mirror technologies for adaptive
		optics and astronomy,} in \enquote{{Adaptive Optics Systems III},} , vol.
	8447 (SPIE, 2012), vol. 8447, pp. 844705--1--18.
	
	\bibitem{Schmid2009}
	J.~H. Schmid, W.~Sinclair, J.~Garc\'{i}a, S.~Janz, J.~Lapointe, D.~Poitras,
	Y.~Li, T.~Mischki, G.~Lopinski, P.~Cheben, A.~Del\^{a}ge, A.~Densmore,
	P.~Waldron, and D.-X. Xu, \enquote{Silicon-on-insulator guided mode resonant
		grating for evanescent field molecular sensing,} Opt. Express \textbf{17},
	18371--18380 (2009).
	
	\bibitem{Fan2009}
	X.~Fan, ed., \emph{{Advanced Photonic Structures for Biological and Chemical
			Detection}} (Springer-Verlag, 2009), 1st ed.
	
	\bibitem{Norte2014}
	R.~A. Norte, \enquote{Nanofabrication for on-chip optical levitation,
		atom-trapping, and superconducting quantum circuits,} Ph.D. thesis,
	California Institute of Technology (2014).
	
	\bibitem{Patent2017}
	\enquote{Method for fabrication of large-aspect-ratio nano-thickness mirrors,}
	Patent pending.
	
	\bibitem{Ji2017}
	X.~Ji, F.~A.~S. Barbosa, S.~P. Roberts, A.~Dutt, J.~Cardenas, Y.~Okawachi,
	A.~Bryant, A.~L. Gaeta, and M.~Lipson, \enquote{Ultra-low-loss on-chip
		resonators with sub-milliwatt parametric oscillation threshold,} Optica
	\textbf{4}, 619--624 (2017).
	
	\bibitem{Born1986}
	M.~Born and E.~Wolf, \emph{Principles of Optics} (Pergamon Press, 1986), 6th
	ed.
	
	\bibitem{Moharam1981}
	M.~G. Moharam and T.~K. Gaylord, \enquote{Rigorous coupled-wave analysis of
		planar-grating diffraction,} J. Opt. Soc. Am. \textbf{71}, 811--818 (1981).
	
	\bibitem{Lu2010a}
	C.~Lu and R.~H. Lipson, \enquote{Interference lithography: a powerful tool for
		fabricating periodic structures,} Laser Photon. Rev. \textbf{4}, 568--580
	(2010).
	
	\bibitem{Starshot2017}
	\enquote{{Breakthrough Initiative -- Starshot},}
	\url{https://breakthroughinitiatives.org/Initiative/3}. Accessed January 2018.
	
	\bibitem{Wiseman1994}
	H.~M. Wiseman, \enquote{Quantum theory of continuous feedback,} Phys. Rev. A
	\textbf{49}, 2133--2150 (1994).
	
	\bibitem{Chakram2014}
	S.~Chakram, Y.~S. Patil, L.~Chang, and M.~Vengalattore, \enquote{Dissipation in
		ultrahigh quality factor SiN membrane resonators,} Phys. Rev. Lett.
	\textbf{112}, 127201 (2014).
	
	\bibitem{Aspelmeyer2014}
	M.~Aspelmeyer, T.~J. Kippenberg, and F.~Marquardt, \enquote{Cavity
		optomechanics,} Rev. Mod. Phys. \textbf{86}, 1391 (2014).
	
	\bibitem{Rhodes1964}
	D.~R. Rhodes, \enquote{On a fundamental principle in the theory of planar
		antennas,} Proc.\ IEEE \textbf{52}, 1013--1021 (1964).
	
	\bibitem{Liu2012}
	V.~Liu and S.~Fan, \enquote{S$^4$:\ A free electromagnetic solver for layered
		periodic structures,} Comput. Phys. Commun. \textbf{183}, 2233--2244 (2012).
	
	\bibitem{LIGO2015}
	{The LIGO Scientific Collaboration}, \enquote{Advanced LIGO,} Class. Quantum
	Grav. \textbf{32}, 074001 (2015).
	
	\bibitem{Saulson1990}
	P.~R. Saulson, \enquote{Thermal noise in mechanical experiments,} Phys. Rev. D
	\textbf{42}, 2437 (1990).
	
	\bibitem{Wilson2012}
	D.~J. Wilson, \enquote{Cavity optomechanics with high-stress silicon nitride
		films,} Ph.D. thesis, California Institute of Technology (2012).
	
	\bibitem{Schmid2011}
	S.~Schmid, K.~D. Jensen, K.~H. Nielsen, and A.~Boisen, \enquote{Damping
		mechanisms in high-Q micro and nanomechanical string resonators,} Phys. Rev.
	B \textbf{84}, 165307 (2011).
	
	\bibitem{Bao2002}
	M.~Bao, H.~Yang, H.~Yin, and Y.~Sun, \enquote{Energy transfer model for
		squeeze-film air damping in low vacuum,} J. Micromech. Microeng. \textbf{12},
	341--346 (2002).
	
\end{thebibliography}
\end{document}